# THE INFLUENCE OF HYDRODYNAMICAL WINDS ON HOT ACCRETION DISK SOLUTIONS

Ranjeev Misra[1]

AND

Ronald E. Taam[1]

Draft version December 16, 2018

## ABSTRACT

The effect of a possible hydrodynamical wind on the nature of hot accretion disk solutions is studied. It is found that the advection dominated branch, in the presence of a wind, maintains the self-similar solution for the disk structure with the temperature, $\theta \propto 1/r$, optical depth $\tau \propto r^{P-1/2}$ and accretion rate $\dot m \propto r^P$. Based on global solutions cooling due to wind energy loss and advection are found to be equally important. For a wide range of viscosity and wind parameters, the temperature is about one-tenth of the virial value and $P \approx 0.9$, independent of the mass accretion rate and radius.

In the context of cooling by unsaturated Comptonization of soft photons, solutions also exist where radiative cooling, advection, and wind cooling are important. In this case, wind regulated solutions are possible. Here, the radial dependence of the critical mass accretion rate above which solutions do not exist is unchanged from those solutions without winds.

The wind/advection dominated solutions are locally unstable to a new type of instability called "wind driven insbility" where the presence of a wind causes the disk to be unstable to long wavelength perturbations of the surface density. The growth rate of this instability is inversely proportional to the ratio of the radiative cooling to the gravitational energy dissipation rates and it can grow on a timescale much longer than the viscous timescale in the disk for sufficiently small radiative cooling efficiencies.

*Subject headings:* accretion, accretion disks-black hole physics -hydrodynamics

## 1. INTRODUCTION

Hot accretion disks are a natural site where externally or internally produced soft photons can be Comptonized to higher energies by disk electrons. It is this property that makes hot accretion disk models attractive in explaining the spectra of compact X-ray binaries and active galactic nuclei. Alternative models to this picture have also been proposed involving shocks (Chakrabarti & Titarchuk 1995), patchy or uniform coronae (Haardt & Maraschi 1993; Liang & Price 1977), or spatially extended regions in a transition disk model (Misra, Chitinis & Melia 1998). Detailed comparisons of the observed X-ray spectra with predictions of these different models can be used to determine the applicability of each model to a given system. In addition, constraints imposed by X-ray timing analysis can also be used to further distinguish between these models (Nowak et al 1999). Apart from these observational tests, theoretical studies of the stability and self-consistencies of these models are important for determining the circumstances under which they are physically realizable.

In a seminal study, Shapiro, Lightman & Eardley (1976) proposed a hot accretion disk model to explain the X-ray spectra of the black hole system Cygnus X-1. Here, a hot inner disk was assumed to be cooled by external soft photons originating from an outer cold disk. In this model, the gravitational energy dissipated in the hot disk heats the protons, and due to the inefficiency of the Coulomb interaction rate in these environments, leads to a large temperature differences between the protons and electrons. The outer cold disk was assumed to be transformed into a hot inner disk for radii less than a critical value due to a thermal instability associated with the dominance of radiation pressure (Lightman & Eardley 1974). Despite the success of this model in explaining the X-ray observations of Cygnus X-1, there remains several outstanding theoretical issues. Specifically, it was not clear whether an unstable cold disk would in reality evolve to a hot disk (Chen & Taam 1994) or whether the narrow transition between the hot and cold disk was energetically possible. Moreover, the hot accretion disk solution, itself, was thermally unstable.

Against this backdrop, the importance of radial advection of energy in such disks was noted by Ichimaru (1977). Taking this effect into account, Narayan & Yi (1994) constructed a second type of hot disk solution with a self-similar structure called Advection Dominated Accretion Flows (ADAF). These solutions were, in general, radiatively inefficient (with most of the energy advected into a black hole) compared to the non-advective, radiatively cooled solutions found earlier. In contrast to the radiatively cooled solutions, ADAFs are thermally stable, and it was shown that a sharp transition between the hot inner regions and the cool outer regions could exist, provided an additional mechanism (e.g. conduction/turbulent diffusion) is assumed which transfers energy from the hot disk to the transition radius (Honma 1996). The topology of these solutions was studied by Chen et al. (1995) and Zdziarski (1998), and it was found that the advection dominated and radiatively cooled solutions only existed below a critical mass accretion rate. Observational support for the radiatively inefficient ADAF solutions was provided by the X-ray spectral fitting of low luminosity galactic centers and quiescent X-ray Novae (Narayan et al. 1998).

[1]Department of Physics and Astronomy, Northwestern University, 2131 Sheridan Road, Evanston,IL 60208; ranjeev@finesse.astro.nwu.edu;taam@apollo.astro.nwu.edu

It was noted by Narayan & Yi (1994) that steady state rotating solutions do not exist for an adiabatic index of the accreting flow, $\Gamma_D = 5/3$. In this case the ADAF solutions become self-similar spherical Bondi accretion flows. This is not a general result, however, since Ogilvie (1999) has shown that time dependent self similar solutions for $\Gamma_D = 5/3$ do exist. If we consider only time independent flows, Narayan & Yi (1994) constructed self-similar rotating solutions in the case of a disk with a nearly equipartition magnetic field (corresponding to an effective adiabatic index of 4/3). Since it is not clear that all accretion flows will contain such fields (e.g., Balbus & Hawley 1998), the applicability of these solutions could be limited. Moreover, Bisnovatyi-Kogan & Lovelace (2000) have recently argued that, in the presence of an equipartition field, the electrons in the disk will be preferentially heated compared to the protons and a two temperature disk may not exist. In contrast, Blandford & Begelman (1999) have recently shown that a rotating self-similar flow with $\Gamma_D = 5/3$ is possible, in the presence of a wind, if the accretion flow rate in the disk varies with radius as $\dot{M} \propto R^P$. A strong hydrodynamic wind from hot disks is possible, in principle, since the ion temperature in the disk is close to the virial value as shown in the early work by Takahara, Rosner, & Kusunose (1989) and Kusunose (1991). Recently confirmation of these results has been obtained by Beckert (2000) for a variety of viscosity prescriptions. In a parallel development, Chakrabarti (1999) and Das (1999) have studied the possibility of outflows in the context of shock/centrifugal barrier models for the accreting flow. Comparisons with observations reveal that the X-ray spectra of such a wind driven self-similar flow called ADIOS (Advection dominated inflow-outflow solutions) can explain the observed spectra of black hole candidates in quiescence (Quataert & Narayan 1999). In addition, Lin, Misra, & Taam (2000) find that the spectral characteristics of high luminosity black hole systems suggest that winds may be important for these systems as well.

Theoretically, the conditions under which wind dominated accreting flows exist are unclear. For example, Abramowicz, Lasota, & Igumenshchev (2000) argue, based on the sign of the Bernoulli function and on detailed two dimensional hydrodynamical simulations that winds are not expected for disks characterized by low viscosities ($\alpha < 0.1$). Although modest outflows have been seen in simulations of disks characterized by high viscosities (Igumenshchev & Abramowicz 1999; see also Stone et al. 1999), the relaxation of the assumption of identical adiabatic indices in the disk and wind (reflecting different mechanisms for driving accretion and outflow) can enhance the effectiveness of winds (see Chakrabarti 1999). The existence of winds can also be affected by the presence of convection which can modify the temperature profile (Narayan, Igumenshchev, & Abramowicz 2000). In particular, they find that the thermal disk structure is dependent on the magnitude of the $\alpha$ viscosity parameter and the direction in which angular momentum is transported by the convection. For outward transport of angular momentum the ADAF solutions are recovered, whereas for inward transport both non-accreting solutions with lowered temperatures ($\alpha < 0.05$) and the standard ADAF solutions are found ($\alpha > 0.05$). Thus the effect of convection on the existence of winds remains inconclusive especially since it is dependent on the unknown distribution of energy dissipation in the vertical extent. It is clear that a strong wind could exist under certain conditions, and it would be useful to study the effects of such a wind on the structure of hot accretion disks.

In ADIOS solutions it is generally assumed that the mass flow rate has a power-law dependence on radius with the power law index, $P$, treated as a parameter (Blandford & Begelman 1999; Beckert 2000). In principle, the variation of the accretion rate with radii should be obtained self-consistently from the surface wind velocity and the specific energy of the wind matter. This, in turn, depends on the disk structure, geometry and the thermal properties of the wind. A realistic computation of this disk/wind configuration is difficult due to the complexity of the problem and the unknown vertical distribution of the viscous dissipation in the disk. However, we show that significant progress can be attained if only the dependence of the surface velocity on the disk structure is estimated and its numerical value is parameterized to take into account the uncertainties involved.

In the next section, the basic equations of hot disks with winds are described. In §3 the results of the calculations are presented and in §4 the stability of the solutions is discussed. Finally, we summarize the results and conclude in the last section.

## 2. DISK STRUCTURE EQUATIONS

The standard equations for a Newtonian, keplerian hot accretion disk are modified to take account of mass, energy, and angular momentum loss due to a hydrodynamical wind. To facilitate the presentation of these equations in a compact form, the following dimensionless quantities are introduced: the ion temperature $\theta \equiv kT_i/m_p c^2$; the electron temperature $\theta_e \equiv kT_e/m_e c^2$; the radius $r = R/r_g$ where $r_g \equiv GM/c^2$ and $M$ is the mass of the black hole; the accretion rate $\dot{m} \equiv \dot{M}/\dot{M}_{Edd}$ where the Eddington accretion rate is defined to be $\dot{M}_{Edd} = 4\pi m_p c r_g/\sigma_T$; Thomson optical depth $\tau \equiv n\sigma_T H$, where $n$ and $H$ are the number density and half-height of the disk respectively; the speed of the wind at the disk surface $v_w \equiv V_w/c_s$ where $c_s = c\theta^{1/2}$ is the local isothermal sound speed of the disk and the Coulomb energy exchange frequency $\nu_c^* = (r_g/c)\nu_c$.

For a non-magnetized two-temperature disk i.e. when the ion temperature ($T_i$) is much higher than the electron temperature ($T_e$), the local pressure is simply $P_g = nkT_i$ and the equation for hydrostatic equilibrium can be written as

$$(H/R)^2 = \theta r \qquad (1)$$

In the presence of a wind the accretion rate is no longer constant and hence mass conservation gives,

$$\frac{d\dot{M}}{dR} = 4\pi R m_p n V_w \qquad (2)$$

which in dimensionless form can be re-written using equ. (1) as,

$$\frac{d\dot{m}}{dr} = \tau v_w r^{-1/2}. \qquad (3)$$

The torque equation (e.g. Beckert 2000) is,

$$\frac{dT}{dR} = \dot{M}(R)\frac{dl}{dR} \quad (4)$$

where $T$ is the viscous torque and where it has been assumed that the specific angular momentum of the wind is same as that in the disk i.e. $l = \sqrt{GMR}$. The viscous integrated stress is described in terms of the $\alpha$-prescription of Shakura & Sunyaev (1973), and it scales as the pressure, i.e. $T = \alpha P_g H$ for a disk in hydrostatic equilibrium. The above equation can then be written in dimensionless form as

$$\alpha \frac{d(\tau \theta r^2)}{dr} = \frac{1}{2}\dot{m}(r)r^{-1/2}. \quad (5)$$

Note that since the accretion rate has a radial dependence (due to the presence of a wind) the above differential equation cannot be trivially integrated.

For a disk in thermal equilibrium, the local vertically integrated energy dissipation rate ($Q_D$) should be equal to the sum of the Coulomb energy loss ($Q_C$), the advective energy flux ($Q_A$) which is non-local radial energy transfer rate associated with the flow and the energy flux carried away by the wind ($Q_W$). The energy dissipation rate due to the viscous torque for each half of the disk is

$$Q_D = \frac{3}{8\pi}\frac{\dot{M}_{Edd}c^2}{r_g^2} \alpha\theta\tau r^{-3/2}. \quad (6)$$

A fraction of this dissipated energy will be transferred by the protons to electrons via Coulomb interactions. The Coulomb energy exchange flux is

$$Q_C = \frac{3}{8\pi}\frac{\dot{M}_{Edd}c^2}{r_g^2} \tau\nu_c^*(\theta - \frac{m_e}{m_p}\theta_e) \quad (7)$$

$$\approx \frac{3}{8\pi}\frac{\dot{M}_{Edd}c^2}{r_g^2} \tau\nu_c^*\theta. \quad (8)$$

Here, the dimensionless Coulomb frequency is given by,

$$\nu_c^* = \frac{r_g}{c}\nu_c = 4.4 \times 10^{-4}\ln\Delta\tau\theta_e^{-3/2}\theta^{-1/2}r^{-3/2} \quad (9)$$

where $\ln\Delta \approx 15$ is the Coulomb logarithm (Spitzer 1962).

The advective energy flux is

$$Q_A = \frac{\dot{M}}{4\pi R^2}\frac{P_g}{nm_p}\xi_A = \frac{1}{4\pi}\frac{\dot{M}_{Edd}c^2}{r_g^2} \dot{m}\theta\xi_A r^{-2} \quad (10)$$

where $\xi_A$ the advective factor is defined to be (Chen et. al. 1995)

$$\xi_A = -(\frac{3r}{2\theta}\frac{d\theta}{dr} - \frac{r}{\tau}\frac{d\tau}{dr} + \frac{r}{H}\frac{dH}{dr}). \quad (11)$$

It has been implicitly assumed in this definition that the specific energy of the gas is $(3/2)kT_i$ or in other words, the adiabatic index of the accretion flow $\Gamma_D = 5/3$. Note that self-similar (ADAF) solutions without winds, $\theta \propto r^{-1}$, $H \propto r$ and $\tau \propto r^{-1/2}$, leads to $\xi_A = 0$. This is consistent and expected since, for $\Gamma_D = 5/3$ and in the absence of a wind, there are no steady self-similar (ADAF) rotating solutions and only spherical Bondi accretion flows are allowed (Narayan and Yi 1994).

The energy flux carried away by the wind is

$$Q_W = nV_W E_W = \frac{1}{8\pi}\frac{\dot{M}_{Edd}c^2}{r_g^2} \tau v_w e_w r^{-5/2} \quad (12)$$

Here $E_W$ is the average energy imparted to a proton in the wind from the local disk and $e_w \equiv E_W/(GMm_p/2R)$ is its dimensionless form. The external energy required for a particle to escape to infinity is $\approx GMm_p/2R - (3/2)(kT_i - kT_f)$ where $T_i$ and $T_f$ are the proton temperature of the wind at the disk surface and at infinity respectively. If the wind is isothermal or if $kT_i << GMm_p/2R$ then this external energy input per particle is simply $GM/2R$ or the binding energy of the particle. Further, if this energy is obtained only from the local region of the disk from where the particle originates, then $e_w \approx 1$. Thus, $e_w < 1$ corresponds to the case where only a fraction of the required energy is obtained from the local region of the disk and the rest presumably is due to heating of the wind from a different region of the disk. The $e_w > 1$ corresponds to the unlikely case where more energy than required is deposited from the local disk to the wind. This additional energy may be radiated away by a radiatively cooled wind and/or appear as kinetic energy of the wind particles at infinity. The uncertainties involved in the energy coupling between the wind and the disk is factored into $e_w$ which is treated as a parameter in this work.

The equation describing conservation of energy ($Q_D = Q_C + Q_A + Q_W$) can be written in a compact form using equs. (6), (8), (10) and (12) as

$$\frac{\nu_c^* r^{3/2}}{\alpha} + \frac{2}{3}\frac{\dot{m}\xi_A}{\alpha\tau r^{1/2}} + \frac{1}{3}\frac{v_w e_w}{\alpha\theta r} = 1. \quad (13)$$

The value of the electron temperature ($\theta_e$) which is required to calculate the Coulomb frequency (equ. 9), depends on the dominant radiative cooling mechanism operating in the hot disk. Following Zdziarski (1998) it is assumed here that the disk cools primarily by unsaturated Comptonization of soft photons. This is motivated by the high energy spectral characteristics of black hole binaries in the low-state since the Comptonization process is characterized by the Compton y-parameter ($y \equiv 4kT_e/m_e c^2$) $\max(\tau, \tau^2) \approx 1$. Thus

$$\theta_e \approx \frac{y}{4max(\tau, \tau^2)} \quad (14)$$

can be used to estimate the electron temperature using $y$ as a parameter of order unity. Technically the electron temperature profile in the disk should be obtained by solving the geometry dependent radiative transfer equations in both vertical and radial directions. Thus it should be emphasized that equation (14) is a simplifying assumption which may be approximately valid only under certain conditions (Shapiro, Lightman & Eardley 1976).

Determining the wind velocity at the surface of the disk is complicated because it depends on the vertical structure of the disk which in turn depends on the detailed variation of the unknown viscosity with height. Furthermore, the opening angle for the wind flow needs to be determined using the geometry of the accreting flow and the pressure gradient in the wind in the radial direction. However, a qualitative estimate of the dependence of the wind velocity with respect to the average temperature and gravitational potential in the disk, can be obtained by studying the simpler geometry of a spherical flow from a radius $R$, temperature $T_i$ and potential $\approx GM/2R$. The factor of two in the potential takes into account the angular momentum of the disk. The estimate obtained from such a spherical flow can be scaled by an unknown factor $\eta$ which would then take into account the uncertainties in the geometry and vertical structure of the disk,

$$v_w = \eta v_s(\theta r) \qquad (15)$$

where $v_s$ is the velocity of the wind at the surface of the sphere divided by the isothermal sound speed ($= \theta^{1/2} c$). Thus the motivation here is to obtain a qualitative dependence of the disk surface wind velocity with temperature and radius of the disk, rather than a quantitative estimation of its magnitude. The standard equations of the hydrodynamical theory of stellar winds have been used to compute the wind velocity at the surface of the sphere. Using Bernoulli's equation and the fact that the Mach number at the sonic point is unity, the following relation for the sonic radius is obtained

$$\frac{r_c}{r_s} = \frac{\lambda[\frac{5-3\Gamma}{4(\Gamma-1)}]}{\frac{v_s^2}{2} + \frac{\Gamma}{\Gamma-1} - \lambda} \qquad (16)$$

where $r_c$ is the location of the sonic point, $r_s$ is the radius of the sphere, $\Gamma$ is the adiabatic index of the wind and $\lambda$ is the ratio of the gravitational potential at the surface to the temperature of the sphere i.e. $\lambda \equiv \phi/kT_i = GMm_p/(2r_s kT_s)$. Note that the adiabatic index in the wind may differ from that in the disk (see Chakrabarti 1999). Conservation of mass and the equation of state, $P_g \propto \rho^\Gamma$ (where $\rho$ is the mass density) gives,

$$v_s = (\frac{\lambda}{2})^{\frac{1}{2}\frac{\Gamma+1}{\Gamma-1}} \; \Gamma^{-\frac{1}{\Gamma-1}} \; (\frac{r_c}{r_s})^{\frac{3\Gamma-5}{2(\Gamma-1)}} \qquad (17)$$

In this work, equations (16) and (17) are solved simultaneously to yield $v_s(\lambda, \Gamma)$ where $\lambda$ for an isothermal disk is $1/(2\theta r)$. If $v_s << \lambda$, then

$$v_s(\lambda, \Gamma) = (\frac{1}{2})^{\Gamma_2} \Gamma^{-\frac{3}{2}} \lambda^2 [(5-3\Gamma)(1-\frac{\lambda(\Gamma-1)}{\Gamma})]^{\Gamma_1} \qquad (18)$$

where $\Gamma_1 = \frac{3\Gamma-5}{2(\Gamma-1)}$ and $\Gamma_2 = \frac{7\Gamma-9}{2(\Gamma-1)}$. For the isothermal case ($\Gamma \to 1$), equ. (18) reduces to

$$v_s(\lambda, \Gamma \to 1) = (\frac{\lambda}{2})^2 \exp(3/2 - \lambda) \qquad (19)$$

There are several points which should be noted here. First, equs. (18) and (19) seem to suggest that $v_s$ has a maxima for variations with $\lambda$. This is an artifact of the assumption $v_s << \lambda$ and if equs. (16) and (17) are used instead (as is done in this work), then $v_s$ is a monotonic function of $\lambda$. Second, there is a maximum value of $\lambda$ beyond which there are no wind solutions or $v_w = 0$ for $\theta < \theta_{max} = (\Gamma - 1)/(2\Gamma r)$. Third, inspection of equ. (18) reveals that for $\theta \gtrsim \theta_{max}$, $v_w$ is a rapidly varying function of $\theta$, a factor two increase in $\theta$ could lead to an increase of two orders of magnitude in $v_w$ (Lin, Misra, & Taam 2000).

The equations corresponding to conservation of mass (3), angular momentum (5) and energy (13) can be solved to obtain the disk structure using the estimated values for electron temperature (14) and wind velocity (15).

## 3. DISK SOLUTIONS WITH WINDS

### 3.1. Local Solutions

The differential equations governing the disk structure can be solved using appropriate boundary conditions. However, these equations can be reduced to algebraic ones, by treating the logarithmic derivatives of physical quantities as constant (i.e., independent of radii) parameters. The solutions for this set of algebraic equations depends only on the local radius and hence are local solutions. Such solutions generally provide important insight on the disk structure and its dependence on disk parameters.

To obtain local solutions it is assumed that the logarithmic derivative of temperature ($\xi_T = -d\log T/d\log R$) and accretion rate ($P = d\log \dot{M}/d\log R$) with respect to the logarithm of the radius are constants. Integrating equation (5) with the constant of integration chosen such the viscous torque vanishes in the inner radius, leads to

$$\tau = \frac{\dot{m}(r) r^{-\frac{3}{2}} J(r)}{(1+2P)\alpha\theta} \qquad (20)$$

where $J(r) \equiv 1 - (6/r)^{P+1/2}$ takes into account the torque free inner boundary condition at $r = 6$. The advective fraction (equ. 11) is reduced to

$$\xi_A = 3(\xi_T - 1) + P \qquad (21)$$

where it has been assumed that $J(r) \approx 1$ is a constant. While it is convenient to consider $\xi_T$ as a parameter, $P$ can be obtained self-consistently by mass conservation (equ. 3),

$$P = \frac{\tau r^{1/2}}{\dot{m}} \eta v_s(\theta r) = \frac{1}{1+2P}(\frac{\eta}{\alpha})(\frac{v_s(\theta r)}{\theta r}). \qquad (22)$$

Hence, the local disk solution is obtained by solving the energy equ. (13) with equs. (20), (21) and (22). Figure 1 shows the variation of the normalized accretion rate, $\dot{m}$, with optical depth, $\tau$, for $r = 40$. Comparison with Figure 1 of Zdziarski (1998) shows that the topology of local solutions does not change with the inclusion of a wind. For accretion rates less than a critical value ($\dot{m} < \dot{m}_{crit}$), there exists a radiatively cooled solution branch and an advection/wind dominated solution branch. No solutions exist for $\dot{m} > \dot{m}_{crit}$. Note, that for $\Gamma = 5/3$, comparisons cannot be made with respect to solutions without winds since steady state solutions do not exist.

For the advection/wind dominated solution branch, the Coulomb exchange rate can be neglected ($Q_C \approx 0$) and equ. (13) can be written for $\xi_T = 1$ as

$$\frac{3}{2} = P(1+2P)\theta r + \frac{e_W}{2}(\frac{\eta}{\alpha})(\frac{v_s(\theta r)}{\theta r})$$
$$= (\frac{\eta}{\alpha}) v_s(\theta r)\bigl(1 + \frac{e_W}{2\theta r}\bigr) \qquad (23)$$

The above with equs. (20) and (22) lead to self-similar solutions for an advection/wind dominated flow: $\theta \propto 1/r$, $\tau \propto r^{P-1/2}$ and $\dot{m} \propto r^P$. Thus, with the inclusion of a wind, self-similar flows are maintained which is a direct consequence of the form of the wind velocity chosen in equ. (15). For example, if the parameter $\eta$ is a function of radius rather than a constant as assumed here, self-similar solutions will no longer be allowed. The solution to equ (23) is shown in Figure 2, where $\theta r$ is plotted as a function of $\eta/\alpha$ for different values of $\Gamma$ and $E_W$. For values of $\eta/\alpha > 2$, $\theta r$ depends only logarithmically with $\eta/\alpha$ and is $\approx 0.1$. The ratio of the wind to advective cooling $Q_W/Q_A = e_W/(2\theta r) \approx 5$ making both cooling channels equally important for this solution branch. It is illustrative to solve for $P$ using equs. (22) and (23) yielding

$$P = \bigl(-1 + \sqrt{1 + 24/(e_W + 2\theta r)}\bigr)/4 \qquad (24)$$

which implies that $P \approx 0.9$ for $e_W \approx 1$ and $\theta r \approx 0.1$. This is shown in Figure 3 where $P$ is plotted as a function of $\eta/\alpha$. Note that from equ. (24) a smaller value of $e_W = 0.5$ corresponds to $P \approx 1.3$ and a larger value $e_W = 1.5$ corresponds to $P \approx 0.7$. Thus P remains of order unity as long as $e_W \approx 1$ which is expected unless the wind is predominately heated non-locally or if the energy transferred by the disk to the wind greatly exceeds the energy required to generate the wind.

At the critical accretion rate, $\dot{m}_{crit}$, radiative cooling becomes comparable to the local dissipation ($Q_C \approx Q_D$). Using the self-similar solution obtained above, $\dot{m}_{crit} \propto r^{3/10}$ which is the same dependence obtained for hot solutions without winds (Zdziarski 1998). The value of $\dot{m}_{crit}$ is however, generally lower than for solutions without winds.

In summary, for a wide range of values for the viscosity parameter ($\alpha$) and the wind parameter ($\eta$), advection/wind dominated solutions are characterized by $\theta \approx 0.1/r$ and $P \approx 0.9$, independent of the accretion rate or radius. We note that the result that the ion temperatures of wind dominated solutions are about an order of magnitude smaller than the no-wind ADAF solutions justifies the assumption of a keplerian disk.

### 3.2. Global Solutions

It is prudent to confirm the inferences obtained from local analysis by constructing global solutions of the disk. These global solutions involve solving the three differential equations for conservation of mass, angular momentum and energy (equs. 3, 5 and 13) with three independent boundary conditions. For convenience, the boundary conditions adopted in this work are that the viscous torque vanishes at the inner radius ($r_i = 6$), and the accretion rate, $\dot{m}(r_o)$, and the temperature, $\theta(r_o)$, are specified at the outer radius, $r_o$. For a given $\dot{m}(r_o)$, $\theta(r_o)$ is computed from the local solution described in §3.1.

Figure 4 illustrates the variation of the mass accretion rate with respect to radius for different values of $\dot{m}(r_o)$. Here, $\theta(r_o)$ is chosen from the advection/wind dominated branch of the local solutions and hence, the global solution presented is also advection/wind dominated. The global solutions are, indeed, similar to the type inferred from local analysis with $P \approx 0.9$ except when $r \gtrsim 6$ where as expected the flow is no longer self-similar. The dotted line is the critical accretion rate (beyond which hot disk solutions no longer exist) computed using local analysis but with values of $P$ and $\xi_T$ taken from the global solution.

Figure 5 presents the global solutions for the case when $\theta(r_o)$ is chosen from the radiatively cooled branch of the local solutions. For $\dot{m}(r_o) \ll \dot{m}_{crit}(r_o)$, the accretion rate is a constant since winds are not important (i.e., $\theta \ll 1/r$). However, for $\dot{m}(r_o) \lesssim \dot{m}_{crit}(r_o)$, winds do become important for certain regions of the disk. In such cases the wind regulates the flow such that $\dot{m}(r) \lesssim \dot{m}_{crit}(r)$. These wind regulated solutions occur for a limited range of $\dot{m}(r_o)$ which is due to the weak dependence of $\dot{m}_{crit}$ with radius ($\dot{m}_{crit} \propto r^{3/10}$). In contrast, the accretion rate at the innermost regions of the disk can be significantly reduced from $\dot{m}(r_o)$ for the advection dominated branch. For radiative cooling mechanisms other than soft photon Comptonization, $\dot{m}_{crit}$ may depend more steeply on radius, leading to a wider range of situations where a wind-regulated accretion flow would exist.

### 4. STABILITY OF DISK SOLUTIONS WITH WINDS

Radiatively cooled hot disks (without winds) are known to be thermally unstable. This result remains valid for the radiatively cooled branch described in §3.1, simply because winds are not important for these solutions. Advection dominated hot disks (without winds) are known to be thermally stable and again this result is not altered for the advection/wind dominated solution since the energy loss rate due to the wind increases with temperature.

Formally the criterion for thermal stability is that the derivative of the energy dissipation rate with respect to temperature should be smaller than the derivative cooling terms with respect to temperature, at a constant surface density. This leads to (Appendix: equ. A1)

$$S_T \equiv \frac{Q_C}{2} + 2Q_A + \xi_{v\theta}Q_W - Q_D > 0 \qquad (25)$$

as the criterion for thermal stability provided that the advection timescale is longer than the thermal timescale. Here, $\xi_{v\theta} \equiv d\log v_s/d\log\theta > 1$. For the wind/advection dominated solutions, $Q_D = Q_A + Q_W$ and $Q_C \ll Q_D$, hence the disk will be locally stable to thermal perturbations. For the radiatively cooled branch, at high accretion rates, advection and wind cooling may be sufficiently important to stabilize the disk.

On timescales longer than the viscous timescale, the surface density (or optical depth) will respond to fluctuations in the disk and its temporal behavior is governed by,

$$\frac{\partial \tau}{\partial t} = -\frac{1}{R}\frac{\partial}{\partial R}(R\tau V_r) - \frac{\tau V_W}{H} \qquad (26)$$

where $V_r$ is the radial velocity. From the torque equation (see equ. 5), it can be seen that fluctuations in $V_r$ are linearly proportional to variations in temperature in the long wavelength limit. On the other hand $V_W/H (\propto v_w)$ is a rapidly increasing function of temperature. Thus, in thermal equilibrium, if $\tau$ is inversely proportional to temperature, the disk will be unstable to long wavelength perturbations. In other words, if a decrease in $\tau$ leads to an increase in the temperature, the mass loss rate by the wind will increase rapidly which will cause a further decrease in $\tau$ leading to a runaway situation.

Technically, for the local wind dominated solution described in §3.1 without radiative cooling ($Q_C = 0$), the temperature of the disk is independent of $\tau$ (equ 23). However, even if there is only a small contribution due to radiative cooling (i.e. $Q_C \ll Q_D$), an increase in $\tau$ will cause $Q_C$ to increase which leads to a decrease in the temperature. Thus, wind dominated solutions are subject to the instability described above in the long wavelength limit. A linear stability analysis leads to the dispersion relation for $\lambda \gg R$ (Appendix: equ. A4)

$$i\omega = \frac{5}{2}\left(\frac{V_r}{R}\right)\left(\frac{Q_C}{S_T}\right)P(\xi_{v\theta} - 1). \qquad (27)$$

For wind dominated solutions, $\xi_{v\theta} > 1$ and the disk is thermally stable ($S_T > 0$), hence $i\omega$ is real and positive indicating that the mode is unstable. For a radiatively cooled disk, $S_T < 0$ and this mode is stable for such solutions. The growth time of the wind instability ($\approx 1/\omega$) can be significantly longer than the viscous (or accretion) timescale ($\approx R/V_r$) if $Q_C/S_T \approx Q_C/Q_D \ll 1$. This implies that the disk structure can be described in terms of local steady state solutions during the linear growth of the wind driven instability. Equ. (27) and the inferred arguments depend on the form of the radiative cooling mechanism (equ. 14) assumed in the analysis. In the general case, $Q_C \propto \tau^\beta$, and it can be shown that the disk will suffer this instability provided $\beta > 1$ and if $S_T > 0$. For the unsaturated Comptonization model (i.e. equ. 14) $\beta = 7/2$ and hence the disk is found to be unstable. Note however, that $\beta < 1$ corresponds to the situation where the electron temperature is proportional to the optical depth and varies faster than $\theta_e \propto \tau^{2/3}$. Since for most physical radiative processes, $\theta_e$ is expected to be inversely proportional to $\tau$, hot disks with winds would in general be unstable. The above results are based on local and linear stability analysis and require confirmation by global simulations.

## 5. SUMMARY AND DISCUSSIONS

The effect of a hydrodynamical wind on a hot accretion disk has been studied for disks characterized by an adiabatic index, $\Gamma_D = 5/3$. It is found that the topology of local ADAF solutions found earlier (for $\Gamma_D < 5/3$) remains the same for $\Gamma_D = 5/3$ with the inclusion of a wind. In particular, there exists two branches of solutions for accretion rates below a critical value, while there are no solutions for higher accretion rates. One solution corresponds to the radiative cooling branch where the effects of the wind are negligible whereas both winds and advection are important for the second branch.

We have shown that the local solutions for the wind/advection dominated branch may be described in a self-similar form with the temperature, $\theta \propto 1/r$, optical depth $\tau \propto r^{P-1/2}$ and accretion rate $\dot{m} \propto r^P$. For a wide range of viscosity and wind parameters, the temperature is about 10% of the virial value ($\theta r \approx 0.1$) and $P \approx 0.9$, independent of the accretion rate and radius. These results were confirmed by global analysis for which self-similar solutions were obtained for radii much greater than the inner disk radius, thereby providing a theoretical basis for the model advanced by Blandford & Begelman (1999). Global solutions also revealed that under certain boundary conditions (see §3.2), the accretion rate $\dot{m} \approx \dot{m}_{crit} \propto r^{3/10}$. For these "critical" solutions radiative cooling and energy loss due to the wind, were both important and hence they are not self-similar.

The wind/advection dominated solutions are thermally stable, but are locally unstable to a new type of instability which we denote as "wind driven instability." The presence of a wind causes the disk to be unstable to long wavelength perturbations in the surface density. Since the growth rate of this instability is inversely proportional to the ratio of the radiative cooling to the gravitational energy dissipation rates, the timescale for the instability could be much larger than the viscous time scale in the disk for a low radiative cooling efficiency.

The accretion rate index, P, $\approx 1$ implies that, if the outer radius of the hot disk is much larger than the inner one, most of the accreting matter is lost in the form of a wind rather than accreted onto the star. If the compact object is a neutron star or white dwarf, it is this residual accretion which gives rise to the X-ray luminosity from the surface of the star. Loeb, Narayan & Raymond (2000) point out that in such a scenario (with nearly similar outer boundary conditions), the surface emission from the star will scale roughly as $\propto R_*^{P-1}$, where $R_*$ is the radius of the star. From the observed luminosities of soft X-ray transients and cataclysmic variables, they estimate $P = 0.9 \pm 0.5$, which is in agreement with the theoretical results presented here.

For this radial dependence of the mass flow rate in the disk, the total energy dissipated in the disk scales as log $(R_o/R_i)$, where $R_o$ and $R_i$ are the outer and inner radii of the hot disk, rather than $R_i^{-1}$. Thus both the inner and outer regions of hot disk are equally important energetically. It is generally assumed that the hot disk is surrounded and connected to an outer cold disk by a narrow transition region. The energy required to make this transition cannot be generated locally if the transition zone is small (Misra & Melia 1996) and is probably due to global transfer from the inner hot regions to the transition zone (Honma 1992, see also Abramowicz et al. 2000). For the wind/advection dominated case this may not be possible since the energy generated in the inner regions is only comparable to that of the outer region. In this case, the transition zone would then have to be extended and could significantly contribute to the luminosity of the source (Misra & Melia 1996). However, the above speculation has to be verified by detailed multi-dimensional simulations of the transition zone taking into account the possibility that the hot disk may be wind dominated.

Since the radiative cooling rate (equ. 8) for the self-similar solution scales as $\propto R^{(14P-15)/4} \approx R^{-1/2}$ for $P \approx 0.9$, most of the luminosity arises from the outer regions of the hot disk. This result, however, depends on the radiative cooling mechanism operating in the hot disk. For example, if the electron temperature is chosen to be $\theta_e \propto R^{\alpha_e}$ instead of equ. 14, then the radiative flux is $\propto R^{2P-3-3\alpha_e/2} \approx R^{-(1+3\alpha_e/2)}$. Hence if $\alpha_e > 2/3$ then the luminosity will be dominated by the inner region.

The terminal velocity of the wind should be approximately equal to the sound speed at the surface of the disk. This implies that the bulk energy per proton of the wind flow at infinity is $\approx 0.1 GM/R$ since the temperature of the disk $\theta \approx 0.1r$ For $P \approx 1$ most of the wind arises from the outermost radius and hence the total bulk energy flux in the wind $\approx 0.1 \dot{M}_{wind} GM/R_o$, where $\dot{M}_{wind}$ is the total mass loss rate and $R_o$ is the outer radius of the hot disk. Since the radiative efficiency of such disks is $<< 0.1$, the bulk energy of the wind flow in some circumstances may be much larger than the radiative flux from the source.

The viscous timescale for the self-similar solution is

$$t_v = \frac{R}{V_R} \approx \frac{R^{1/2}}{(1+2P)(\theta r)\alpha} \left(\frac{r_g}{c}\right) \qquad (28)$$

which numerically is $t_v \approx 250 (r_g/c) \approx 0.01$ s for $\alpha = 0.1$, $R = 50 r_g$ and a 10 $M_\odot$ black hole. Thus, a wind driven instability may occur on timescales $\sim 1 - 10$ s for $Q_C/Q_D \approx 0.1 - 0.01$. The instability may not lead to disruption of the disk, but instead give rise to large amplitude oscillations. For example, the surface density could increase, accompanied by a decreasing temperature. When the temperature decreases sufficiently, the wind is quenched and radiative cooling becomes important. Since the radiatively cooled disk will be thermally unstable, the system may return to the wind dominated state on thermal timescales, giving rise to cyclic behavior. Alternatively, the surface density may decrease and the disk may then be evacuated and finally filled again on the viscous timescale. In either case, the disk may evolve through epochs of high wind activities on timescale ≈ seconds. Such a disk/wind interaction may be important for explaining the spectral and temporal behavior of the regular bursting activity exhibited by the microquasar GRS 1915-105 (Taam et al 1997).

Future work involving time dependent numerical simulations of hot disk with winds are needed to confirm the above speculations and to quantify the rate of mass loss and varying spectral characteristics during such events. Such studies may also shed light on the relevance of wind instabilities as a mechanism for either oscillatory behavior or evaporative behavior in the hot inner disk regions. Besides these issues, further theoretical investigations of the transition from a cold to a hot wind dominated disk and the role of two dimensional effects along the lines suggested by Turolla & Dullemond (2000) would also be illuminating.

R.M. gratefully acknowledges support from the Lindheimer Fellowship at Northwestern University.

APPENDIX

LINEAR STABILITY ANALYSIS IN THE LONG WAVE-LENGTH APPROXIMATION

The linear response of the disk structure is calculated for perturbation of the ion temperature $\theta \to \theta(1 + \delta\theta e^{i(\omega t - kr)})$ and optical depth $\tau \to \tau(1 + \delta\tau e^{i(\omega t - kr)})$, where $\delta\theta$ and $\delta\tau$ are small dimensionless quantities. In the long wavelength approximation, it is assumed that $kr << 1$ such that $e^{ikr} \approx 1$ which implies that the radial derivatives of perturbed quantities can be neglected.

In this formalism, the linear response of physical quantities (e.g., the energy flux) can be obtained directly from the dependence of the quantity on temperature and optical depth. The dissipative energy rate $Q_D \propto \theta\tau$ (equ. 6) and hence $\delta Q_D = \delta\theta + \delta\tau$. The Coulomb energy exchange flux $Q_C \propto \tau^{7/2}\theta^{1/2}$ (equs. 8, 9, 14) while the energy flux carried away by the wind $Q_W \propto \tau v_w(\theta)$ (equ. 12 ). The advective energy flux (equ. 10), $Q_A \propto \dot{m}\theta \propto \tau\theta^2$ since $\dot{m} \propto \tau\theta$ (equ. 5) for a self-similar steady state solution with $kr << 1$. Note that in the long wavelength approximation the advective factor ($\xi_A$) is invariant to perturbations. Thus the normalized response of the total energy flux $Q_T \equiv Q_D - Q_C - Q_W - Q_A$ is,

$$Q_T\delta Q_T = (Q_D - \frac{1}{2}Q_C - \xi_{v\theta}Q_W - 2Q_A)\delta\theta + (Q_D - \frac{7}{2}Q_C - Q_W - Q_A)\delta\tau$$
$$= -S_T\delta\theta - \frac{5}{2}Q_C\delta\tau \qquad (A1)$$

where $\xi_{v\theta} \equiv d\log v_s/d\log\theta$. On thermal timescales $\delta\tau << \delta\theta$ and $-Q_T\delta Q_T/\delta\theta = S_T > 0$ is the criterion for thermal stability (equ. 25).

The time dependent form of the mass conservation equation (3) governs the temporal behavior of the surface density or optical depth (equ. 26)

$$(\frac{r_g}{c})\frac{\partial\tau}{\partial t} = \frac{1}{r}\frac{\partial\dot{m}}{\partial r} - \tau v_w r^{-3/2} \qquad (A2)$$

Noting that $\dot{m} \propto \tau\theta$, the linearized form of the above equation is

$$i\omega\tau\delta\tau = (\frac{c}{r_g})(\frac{1}{r}\frac{\partial\dot{m}}{\partial r})\delta\theta(1 - \xi_{v\theta}) \qquad (A3)$$

For timescales much longer than the thermal timescale, $\delta Q_T \approx 0$ i.e. the system will be in thermal equilibrium. Combining equations (A1) and (A3), yields the dispersion relation in the long wavelength limit (equ. 27)

$$i\omega = \frac{5}{2}(\frac{V_r}{R})(\frac{Q_C}{S_T})P(\xi_{v\theta} - 1) \qquad (A4)$$

where $V_r$ is the magnitude of the radial velocity at radius $R$.

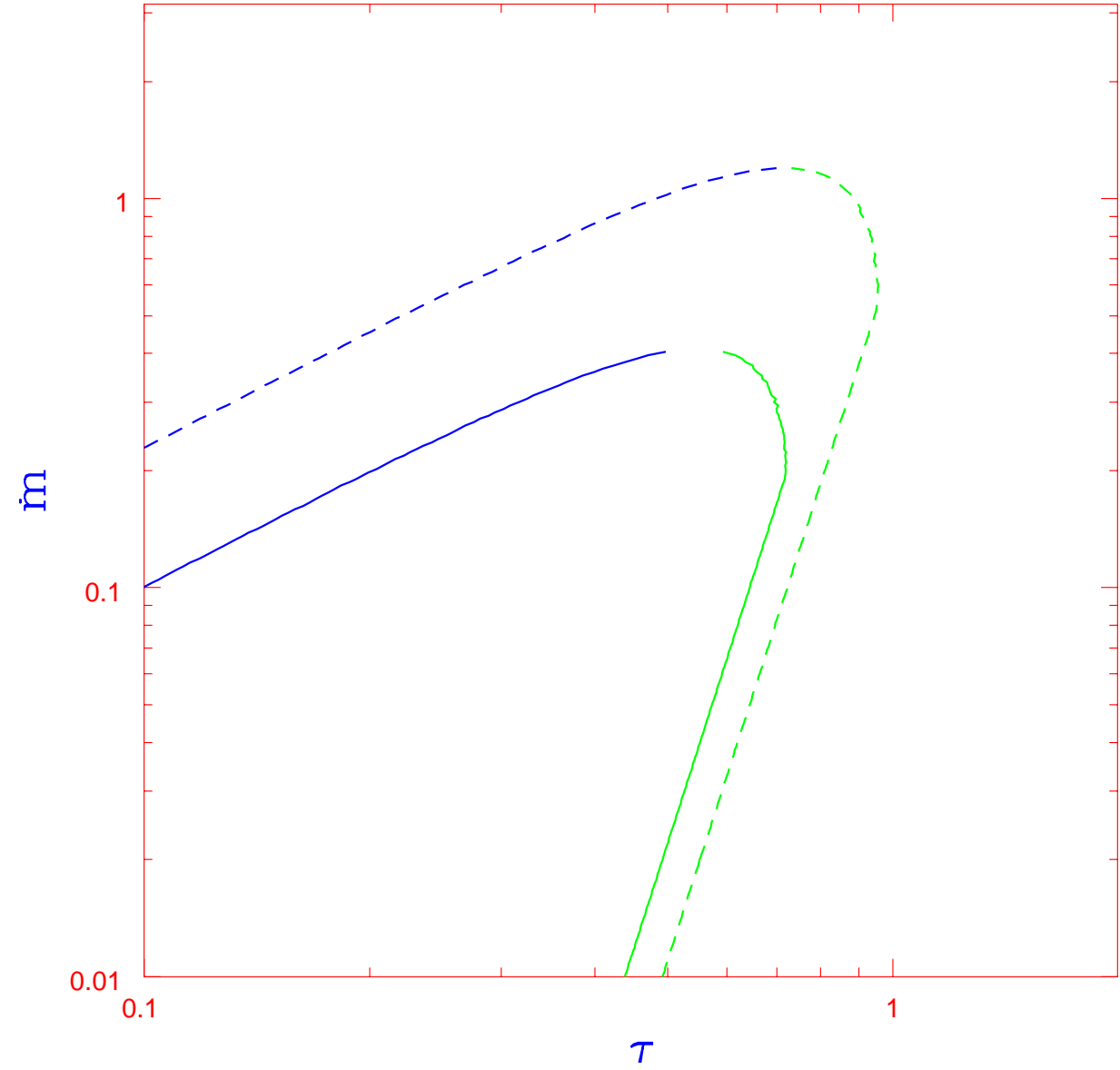

Fig. 1.— Variation of the normalized accretion rate ($\dot{m}$) with optical depth ($\tau$) for $r = 40$. Solid line: $\alpha = 0.5$; Dashed line: $\alpha = 1.0$. Other parameters are $\eta = 1, e_W = 1, y = 1, \Gamma = 1.2$ and $\xi_T = 1$.

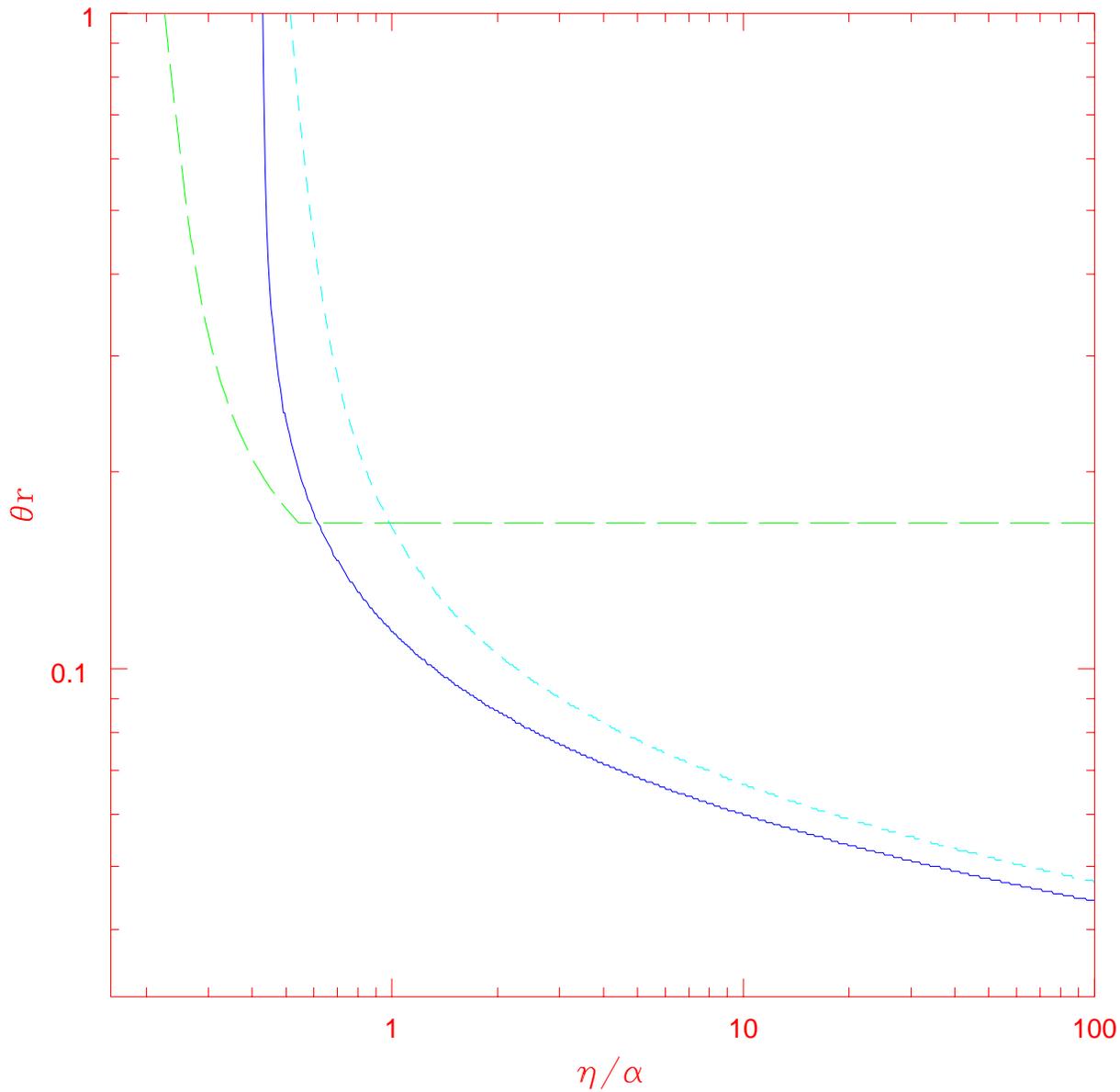

Fig. 2.— Variation of $\theta r$ with $\eta/\alpha$ for $\Gamma = 1.01$, $e_W = 1$ (solid line), $\Gamma = 1.01$, $e_W = 0.5$ (short dashed line) and $\Gamma = 1.2$, $e_W = 1$ (long dashed line).

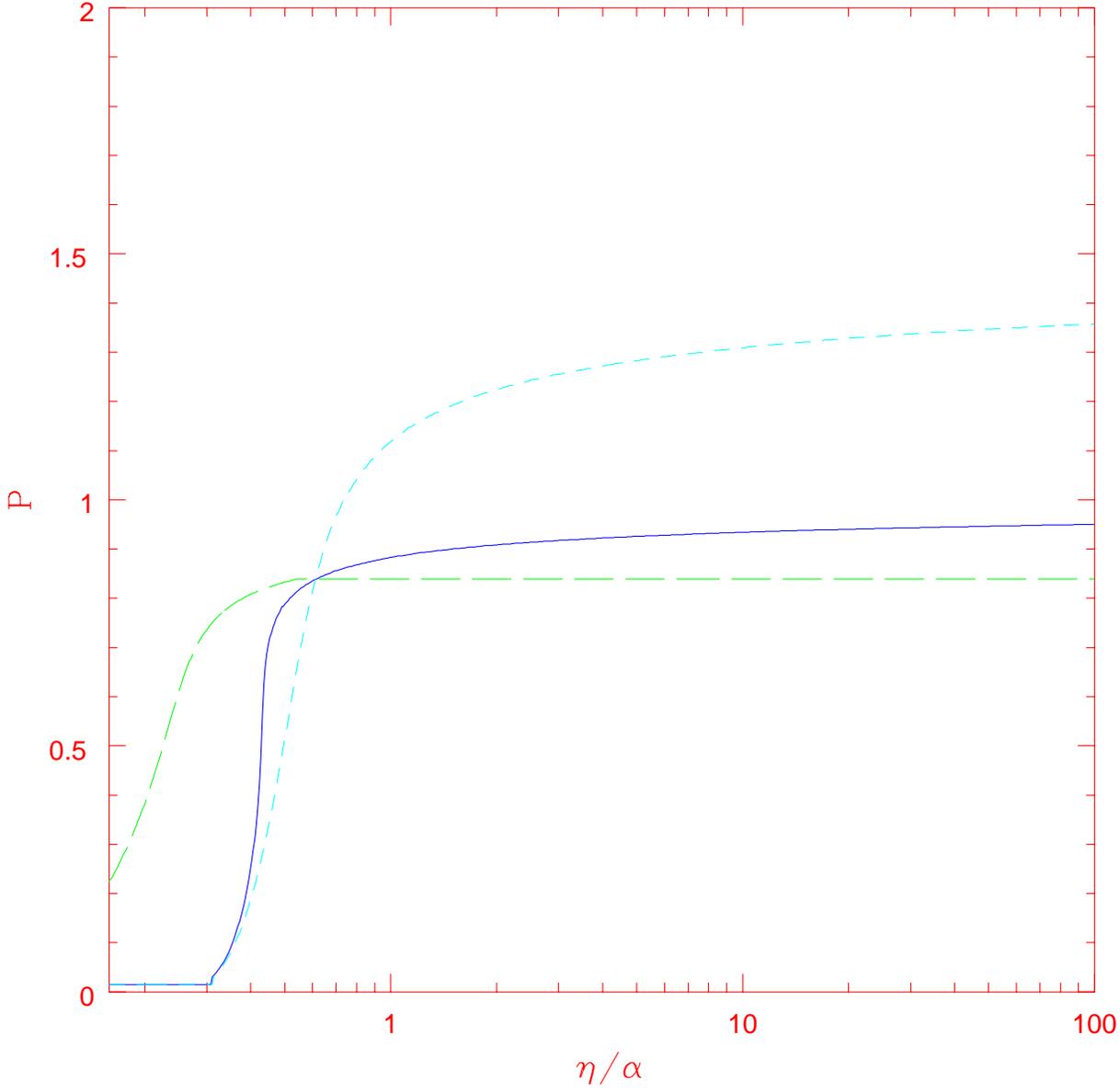

Fig. 3.— Variation of the accretion slope $P$ with $\eta/\alpha$ for $\Gamma = 1.01$, $e_W = 1$ (solid line), $\Gamma = 1.01$, $e_W = 0.5$ (short dashed line) and $\Gamma = 1.2$, $e_W = 1$ (long dashed line).

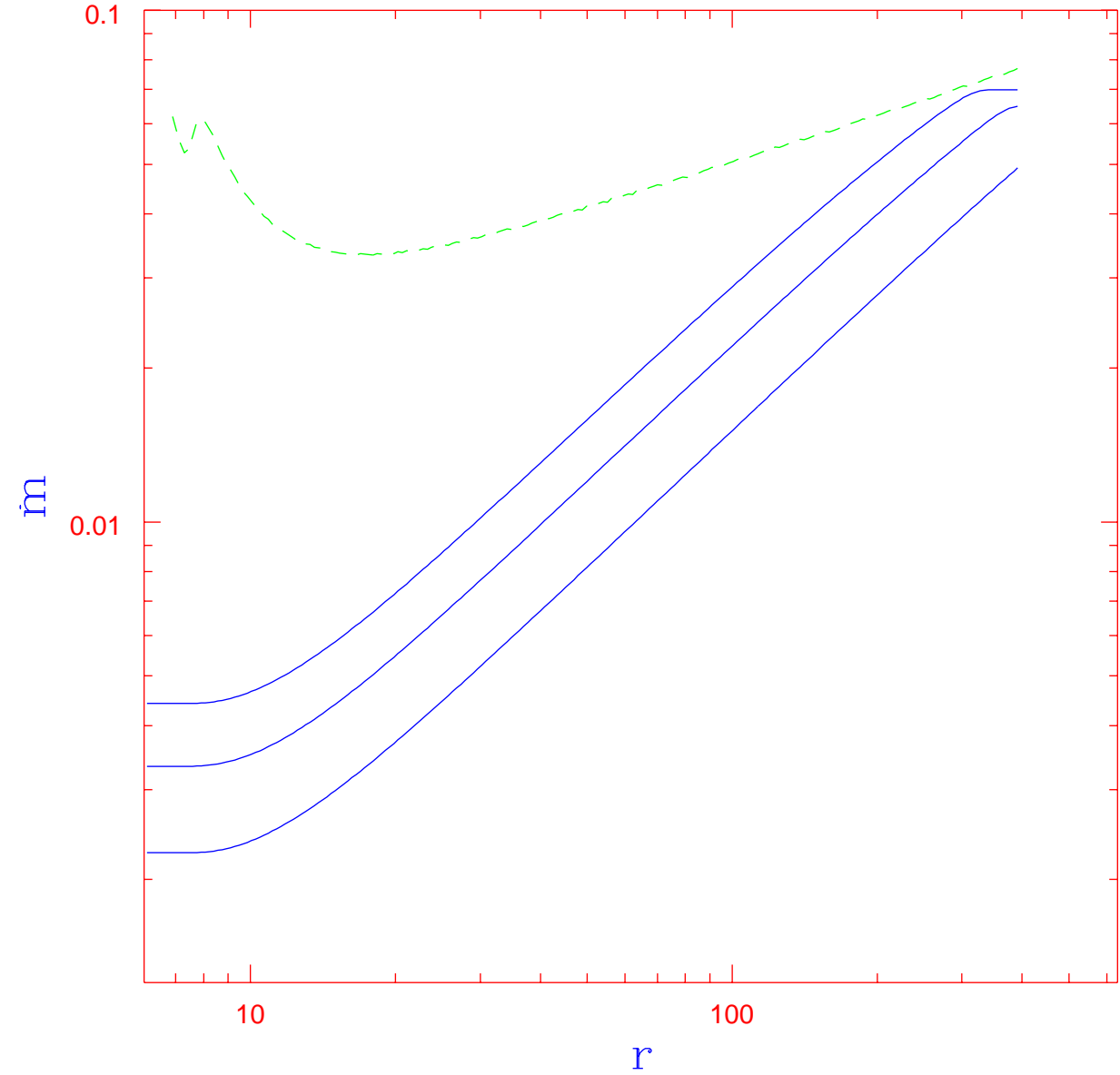

Fig. 4.— Accretion rate vs. radii for different values of accretion rate at the outer boundary ($r = 400$). The temperature at the outer boundary is chosen from the advection/wind dominated branch of the local solutions. Dashed curve represents the local critical accretion rate. Parameters used for the plot are $\alpha = 0.1$, $y = 1$, $\eta = 1$, $e_W = 1$ and $\Gamma = 1.2$

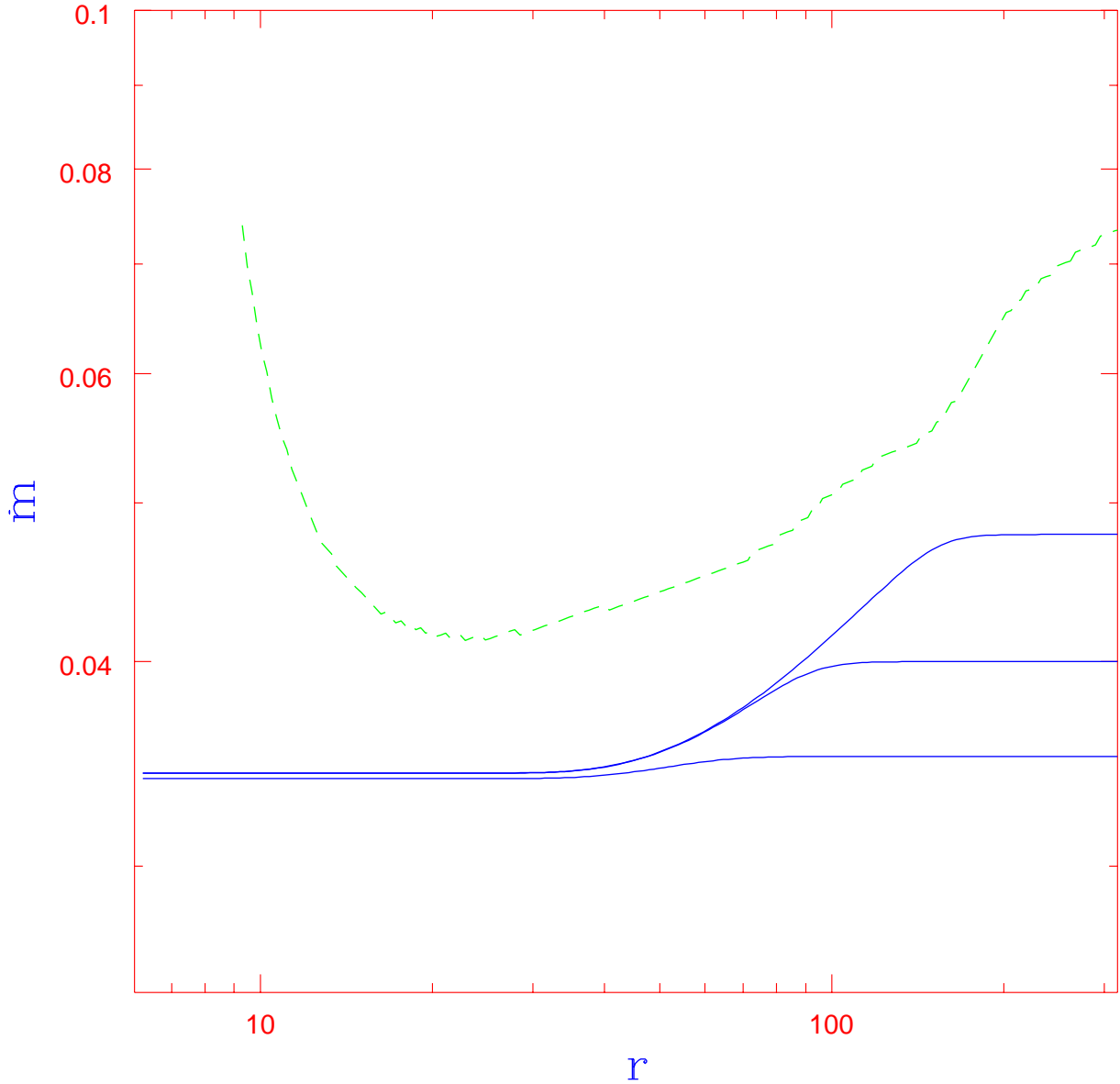

Fig. 5.— Accretion rate vs. radii for different values of accretion rate at the outer boundary ($r = 400$). The temperature at the outer boundary is chosen from the radiatively cooled branch of the local solutions. Dashed curve represents the local critical accretion rate. Parameters used for the plot are $\alpha = 0.1$, $y = 1$, $\eta = 1$, $e_W = 1$ and $\Gamma = 1.2$